\documentstyle[epsf,sprocl]{article}
\def\be{\begin{equation}}
\def\ee{\end{equation}}
\def\bea{\begin{eqnarray}}
\def\eea{\end{eqnarray}}

\begin{document}

\title{ Mobility of dislocations in semiconductors }

\author{K. Stokbro}

\address{Mikroelektronik Centret, Technical University of Denmark, 
DK-2800 Lyngby, Denmark}

\author{L. B. Hansen}

\address{Center for Atomic-scale Materials Physics(CAMP), Technical
University of Denmark, DK-2800 Lyngby, Denmark}

\author{ B. I. Lundqvist }

\address{Department of Applied Physics, Chalmers University of
Technology and Goteborg University, S-41296 Goteborg, Sweden}

\maketitle\abstracts{Atomic-scale calculations for the dynamics of the
90$^0$ partial glide dislocation in silicon are made using the
effective-medium tight-binding theory. Kink formation and migration
energies for the reconstructed partial dislocation are compared with
experimental results for the mobility of this dislocation. The results
confirm the theory that the partial moves in the dissociated state via
the formation of stable kinks. The correlation between glide
activation energy and band gap in semiconducting systems is
discussed.}

Metallic systems  are often ductile even at low temperatures, which is
due to the high mobility of dislocations in these systems. In
semi-conducting systems, on the other hand, there are large Peierls barriers which must be overcome
in order to move a dislocation and these materials therefore behave
brittle.
The low mobility of the dislocations is due to the electronic structure in
the semiconducting systems, and a   direct proportionality between the
band gap and the dislocation glide activation energy has been
observed.\cite{Gi75} Another difference between the metallic 
and the semiconducting systems, is that in the diamond cubic lattice
there are  two distinct (1,1,1) glide planes, giving rise to
two different sets of dislocations, called the glide set and the
shuffle set. Figure~1 shows the position of the two different 
slip planes, and we see that while the glide plane breaks three 
nearest neighbour bonds per atom the shuffle plane only breaks 
one nearest neighbour bond per atom. 

In this paper we will address the
problem of which slip plane is important for dislocation glide in semiconductors.
Our interest for this problem arose from a paper by
Gilman,\cite{Gi93}  where he based on the empirical relation between 
the dislocation glide activation energy and the band gap, 
argues
   that the shuffle set is the
relevant dislocations in semiconductors. However, this seems to contradict
high resolution transmission electron microscopy images of dislocations in 
silicon, since these show dissociated dislocations and this
 is  only consistent with models of glide dislocations. 

\begin{figure}[t]
\vspace*{-2.5cm}
\mbox{\hspace{-.5cm}
\epsfxsize=12.0cm
\epsfbox{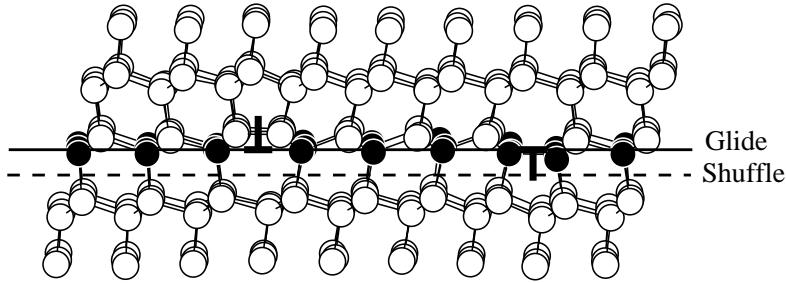}
}

\vspace*{-4.0cm}

\caption{Sideview of the $8\times9\times3$ atom unit cell used for the
EMTB calculation. The horizontal solid line shows the glide plane of the
two partials and their positions are marked along the line. The dashed
line shows the position of the shuffle plane.}
\end{figure}

To  investigate whether the mobilities of the glide dislocations are in
accordance with experimentally observed dislocation mobilities in
semiconductors, 
we have used the
effective-medium tight-binding
model(EMTB)\cite{StChJaNo94b}
to calculate the mobility of a particular glide dislocation,
the 90$^0$ edge dislocation in silicon. The EMTB model is a total
energy tight-binding  method
based on Effective-Medium Theory\cite{JaNoPu87} using a first-order
Linear Muffin-Tin Orbital(LMTO)
tight-binding model\cite{AnJeSo87} to calculate the band-structure energy.
 The model gives a quantum mechanical description 
of silicon and  previous studies  have demonstrated the
ability of the model to accurately describe dislocations in
silicon.\cite{HaStLuJaDe95} For the atomic simulation we have used a 
 $8\times9\times3$ atom unit cell containing two 90$^0$ partial edge
dislocations with opposite Burgers vector (b=2.18\AA ), such that
periodic boundary conditions can be used. Figure~1 shows a sideview of
the unit cell, and Fig.~2(a) shows a top view  of one of the partials. 
Note that the atoms in the core region(along the solid line) relax
asymmetrically to obtain fourfold coordination. We find that  this structure
has an energy that is 0.18 eV/\AA \ lower than a symmetric arrangement,
where the atoms have ``quasi fivefold'' coordination,   in good
agreement with the {\it ab initio}
result by Bigger {\it et. al.}\cite{BiMcSuPaStKiBiCl92} of 0.2 eV/\AA .

Dislocation glide is believed to proceed via the formation of stable
kink pairs and their subsequent spreading along the dislocation
line.\cite{HiLo82} At low stresses the velocity $v_{d}$ of the steady 
state motion of the dislocation is in this theory given by
\begin{equation}
v_{d} \propto \exp [-(U_{dk}/2 + W_m)/k_bT],
\end{equation}
where  $U_{dk}$ is the formation energy for a  double kink, and $W_m$
is the migration energy.

\begin{figure}[t]
\mbox{\hspace{-1cm}
\epsfxsize=14.0cm
\epsfbox{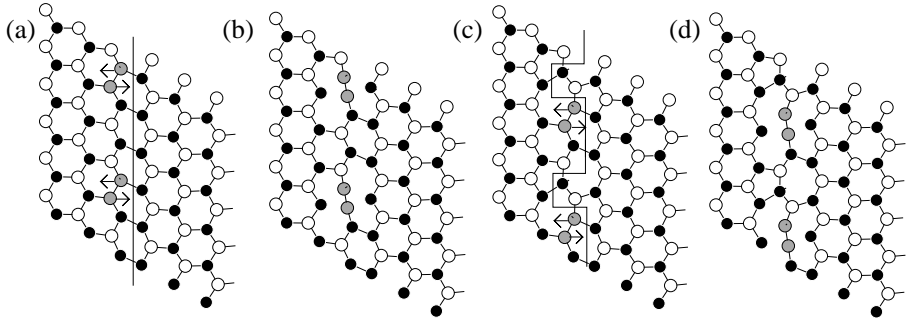}}
\vspace{-1.5cm}
\caption{ (a) The perfect reconstructed dislocation.  Dashed atoms
are moved to create a kink pair. 
(b) Transition state for forming a  kink pair.  
(c) Kink pair of separation $3.8$ \AA .
(d) Transition state for   kink migration. The unit cell
is repeated twice in the vertical direction, and only part of the unit
cell is shown in the horizontal direction.\label{fig:kink}}
\end{figure}

To  find these energies we first create a kink by moving the shaded
atoms in Fig.~2(a) in the direction of the arrows. The reaction
coordinate is chosen as the difference between the coordinates of the 
two moving atoms in the direction of the arrows, and all other degrees
of freedom in the unit cell is  allowed to relax. The double
kink of separation $x=3.8$\AA \
is shown in Fig. 2(c), and we find the formation energy to be
$U_{dk}(3.8 {\rm \AA})=0.3$ eV.  Standard elasticity
theory\cite{HiLo82} predicts   an elastic attraction
between the kinks of $\approx
0.1$ eV at this separation,   and  we therefore  estimate $U_{dk}=0.4$
eV. Next we calculate the kink migration barrier, using the  reaction
path indicated by the arrows
 in Fig. 2(c). 
Figure 2(d) shows the transition state, and from the energy 
 we find $W_m = 1.45$ eV.  The best experimental
estimates\cite{GoHiSaSpAl93,FaIuNi86} give values $U_{dk}\approx 1.0$
eV and $W_m \approx 1.6$ eV, and another tight-binding calculation by
Nunes {et. al.}\cite{NuBeVa96} have found the values $U_{dk}\approx1.0$ eV
and $W_m \approx 1.8$ eV. Our calculated migration barrier is in good
agreement with these values, while the kink
formation energy is too low. We believe that the latter is due to a
strong attraction between the two kinks, which due to their close
proximity is not described by  elasticity theory.
We also note that  the atomic structure  in the transition
state is very similar to the  symmetric reconstruction of the
core structure, and from this analogy we estimate  a barrier of
$7.6 \; {\rm \AA } \times 0.18 \; {\rm eV/\AA } = 1.5$ eV in good agreement with the actual calculated value.

If we use the experimental value for the kink formation energy,
we find an effective barrier for dislocation glide
of $U_{eff} = U_{dk}/2 + W_m= 2.0$ eV, nearly  twice
the silicon bandgap ($E_g=1.1$ eV\cite{Ph73}). Since the same relation
is found for other semiconductors it is tempting to search for a
simple explanation for this proportionality. By inspecting our
calculation we find that the main contribution to the migration
barrier is from the one-electron term, i.e. the  barrier is  due to
the electronic structure in the transition state. At the transition
state there are states in the band gap(the same is
the case for the symmetric reconstruction of the
core structure,\cite{BiMcSuPaStKiBiCl92}) and  we therefore expect
 a proportionality between the bandgap and the migration barrier.
Work is in progress to make a more quantitative model of the relation
between the dislocation activation energy and bandgap of
semiconductors.\cite{lars} 

In conclusion we have used the EMTB model to calculate the effective
barrier for dislocation glide of the 90$^0$ partial in silicon. The
calculated barrier is in good agreement with experimental findings,
and we therefore conclude that the glide set is the relevant
dislocations in silicon. Furthermore,  inspection of the different
contributions to the total energy suggests  a 
 proportionality between the semiconductor bandgap and the migration
barrier of this dislocation, and the
results might therefore have
implications for other semiconductors as well.

\end{document}